\documentclass[useAMS,usenatbib]{mn2e}
\usepackage{graphicx}
\def\rmscr#1{{\hbox{\rm \scriptsize #1}}}

\begin{document}
\title[Diffractive Microlensing III]{Diffractive Microlensing III:
  Astrometric Signatures}
\date{Accepted 2010 October 4.  Received 2010 September 30; in
  original form 2010 March 1}
\author[J. S. Heyl]{Jeremy S. Heyl\\
Department of Physics and Astronomy, University of British Columbia,
Vancouver, British Columbia, Canada, V6T 1Z1;\\
 Email: heyl@phas.ubc.ca; Canada Research Chair}
\pagerange{\pageref{firstpage}--\pageref{lastpage}} \pubyear{2009}

\maketitle

\label{firstpage}

\begin{abstract}
  Gravitational lensing is generally treated in the geometric optics
  limit; however, when the wavelength of the radiation approaches or
  exceeds the Schwarzschild radius of the lens, diffraction becomes
  important.  Although the magnification generated by diffractive
  gravitational lensing is well understood, the astrometric signatures
  of diffractive microlensing are first derived in this paper along
  with a simple closed-form bound for the astrometric shift.  This
  simple bound yields the maximal shifts for substellar lenses in
  solar neighbourhood observed at 20~GHz, accessible to
  high sensitivity, high angular resolution radio telescopes such as
  the proposed Square Kilometre Array (SKA).
%   The development of the Square Kilometre Array (SKA) will open a new
%   window on the Universe.  In particular the SKA will combine
%   unprecedented sensitivity with high angular resolution.  This
%   combination may allow the detection of astrometric signatures from
%   microlensing events by nearby objects against more distant radio
%   sources --- the sources of interest in this case are quasars.
%   Additionally the long wavelength of the radiation (radio versus
%   optical) may also allow the detection of diffractive microlensing
%   that often amplifies the astrometric signature.  An astrometric
%   monitoring campaign either with the SKA or a purpose-build
%   lower-sensitivity array is proposed.
\end{abstract}

\section{Introduction}

Gravitational microlensing is a powerful tool to probe the
constituents of the solar neighbourhood, the Galaxy and beyond
\citep[e.g.][]{wambsganss06:_gravit_lensin}.  In particular
\citet{2005ApJ...635..711G} have propose astrometric microlensing as a
technique to detect sub-stellar objects in the solar neighbourhood,
and \citet{2009arXiv0910.3922H,2010arXiv1002.3007H} argued that
diffraction could provide important constraints on lensing objects in
the Kuiper belt and beyond.   The combination of diffraction and
astrometric lensing offers a new dimension to microlensing surveys.  

Several authors have examined gravitational lensing including the
effects of diffraction
\citep[e.g.][]{1978RaF....21...87I,1980Ap&SS..71..171E,1981Ap&SS..78..199B,1986ApJ...307...30D,1995ApJ...442...67U,2004A&A...423..787T}.
However, the focus has almost entirely been on the magnification of
the image.  An exception is the work of \citet{Labeyrie:1994p1906}
that examines the possibility of using a planetary mass lens as a
telescope.   This letter will examine the astrometry of diffractive
lensing; that is how does lensing affect the centroid of the light
distribution including the effects of diffraction.   As diffraction
can amplify the magnification of a gravitational lens, so too does it
increase the motion of the image.   Measuring the motion of the image
can provide constraints on the lens, source and their relative motion.

The commissioning of the Square Kilometre Array (SKA) over the next
decade will offer an unprecedented view of the radio sky.
\citet{2000pras.conf..213K} outlines some prospects for using the SKA
to understand strongly lensed quasars and especially the small-scale
structure of the lensing object.  This letter also examines primarily
the lensing of quasars but focuses on nearby lensing objects with the
hopes to provide constraints on the number of small bodies in the
solar neighbourhood.  Such constraints are difficult to obtain
otherwise.  The letter is divided into a calculation
(\S~\ref{sec:calculations}) of the astrometric signature of lensing
both in the diffractive and geometric optics regimes, a description of
the results (\S~\ref{sec:results}) and an evaluation of the prospects
of observing this effect (\S~\ref{sec:conclusions}).

\section{Calculations}
\label{sec:calculations}
\citet{1992grle.book.....S}  give the magnification for a point source
including diffraction
\begin{equation}
\mu_\omega =  \left | I \right |^2
= \left | \int_{u_d}^\infty u^{1-if} e^{iu^2/2} J_0(uv) du \right |^2.
\label{eq:1}
\end{equation}
where $u$ is the radial coordinate that integrates over the plane of
the lens, $J_0$ is the Bessel function of the first kind, and $v$ is the impact
parameter of the source relative to the lens.  Both $u$ and $v$ are
dimensionless and measure lengths in units of the reduced Fresnel
length,
\begin{equation}
l_\rmscr{Fr} =\sqrt{\frac{c}{\omega_d} \frac{D_d
    D_{ds}}{D_s}}
\end{equation}
where $\omega_d=\omega(1+z_d)$ is the angular frequency of the
radiation as it passes the lens.  Hence the value of $u_d$ which
compares the angular size ($r_d/D_d$) of the occulting portion of the
lens to angular scale of its diffraction pattern ($\lambda/r_d$) is
given by
\begin{equation}
u_d = r_d \sqrt{\frac{\omega_d}{c} \frac{D_s}{D_d
    D_{ds}}} = \frac{r_d}{l_\rmscr{Fr}}. \label{eq:6}
\end{equation}
The parameter $f$ is given by 
\begin{equation}
f=\frac{\omega_d}{c} \frac{D_s}{D_d D_{ds}} R_E^2
\label{eq:2}
\end{equation}
where $D_s$ is the distance to the source, $D_d$ is the distance to
the lens, $D_{ds}$ is the distance between the source and the lens,
$z_d$ is the redshift of the lens, and 
the Einstein radius is the characteristic length of the lens,
\begin{equation}
R_E = \sqrt{2 R_S \frac{D_d D_{ds}}{D_s} } = \sqrt{f} l_\rmscr{Fr} \label{eq:3}
\end{equation}
for the Schwarzschild lens where $R_S=2 G M_d/c^2$ where 
$G$ is Newton's gravitational constant, $M_d$ is the mass of the lens
and $c$ is the speed of light; therefore, the value of 
\begin{equation}
f = 2 R_S \frac{\omega_d}{c} \label{eq:4}.
\end{equation}
The parameter $f$ compares the wavelength of the radiation to the
Schwarzschild radius of the lens. The limit where the gravitational
field of the lens is negligible is $f=0$, so the effect of gravity on
the form of the integral, Eq.~(\ref{eq:1}), is quite modest.  In a
cosmological context all of the distances given are angular diameter
distances.

Here the occultation will be neglected ({\em i.e.}\ $u_d \rightarrow 0$). The
integral can be calculated in closed form in terms of the confluent
hypergeometric function ($_1F_1(a;b;z)$) for $u_d=0$.  \citet{Grad94}
give relation (6.631.1) which in this particular case yields
\begin{eqnarray}
\int_0^\infty u^{1-if} e^{iu^2/2} J_0(uv) du \!\!&=&\!\!e^{\pi f/4} 
e^{i \left ( \pi - f \ln 2\right)/2}
\Gamma \left (1 - i \frac{f}{2}\right ) \times \nonumber \\
& & ~{}_1F_1 \left ( 1 - i \frac{f}{2}; 1 ; -i\frac{v^2}{2} \right )\label{eq:5}.
\end{eqnarray}
The result for $f=0$ is simply $i\exp(-iv^2/2)$.

\subsection{Astrometry}
   
The gradient of the phase of the incoming radiation points to the
apparent location of an unresolved source on the sky.  This location
on the image plane is given by
\begin{equation}
{\bar u} = -\Im \frac{\partial \ln I}{\partial v}
= \Im \left [ \frac{1}{I} \int_{u_d}^\infty u^{2-if} e^{iu^2/2}
  J_1(uv) du \right ].
\label{eq:8}
\end{equation}
where $\Im$ denotes the imaginary part of the expression it precedes.
The first equality will also hold for an asymmetric lens where $\langle
\vec u \rangle = -\Im \nabla \ln V$.
If $u_d=0$ the following expression holds
\begin{equation}
{\bar u} = v \Re \left [   \left ( 1 - i \frac{f}{2} \right )
  \frac{{}_1F_1 \left ( 2 - i \frac{f}{2}; 2 ; -i\frac{v^2}{2} \right
    ) }{{}_1F_1 \left ( 1 - i \frac{f}{2}; 1 ; -i\frac{v^2}{2} \right
    ) } \right ]
\label{eq:9}
\end{equation}
where for $f=0$ the quantity in the brackets is unity and $\Re$
denotes the real part of the expression in brackets.

For values of $f, v\ll 1$ the ratio of the hypergeometric functions can
be conveniently approximated by Gauss's continued fraction
\citep{Cuyt08}
% \def\cfrac#1#2{\displaystyle\frac{#1}{#2}}
% \begin{equation}
% \frac{{}_1F_1(a;b+1;z)}{b{}_1F_1(a;b;z)} = \cfrac{1}{b + \cfrac{a
%     z}{(b+1) + \cfrac{(a-b-1) z}{(b+2) + \cfrac{(a+1) z}{(b+3) +
%         \cfrac{(a-b-2) z}{(b+4) + {}\ddots}}}}}
% \end{equation}
\begin{equation}
  \frac{{}_1F_1 \left ( a+ 1; b+1 ; z \right )}
    {{}_1F_1 \left ( a ; b; z \right )}
      =\displaystyle\frac{1}{1 - \displaystyle\frac{{\displaystyle\frac{b-a}{b(b+1)}z}}
{1 + \displaystyle\frac{{\displaystyle\frac{a+1}{(b+1)(b+2)}z}}
{1 - \displaystyle\frac{{\displaystyle\frac{b-a+1}{(b+2)(b+3)}z}}
{1 + \displaystyle\frac{{\displaystyle\frac{a+2}{(b+3)(b+4)}z}}{1 - \ddots}}}}}.
\end{equation}
% The confluent hypergeometric function satisfies Kummer's equation
% \citep{Abro70}.  The differential equation can be arranged as 
% \begin{equation}
% z \left [ \frac{d}{dz} \left ( \frac{w'}{w} \right ) + \left (
%     \frac{w'}{w} \right )^2 \right ] + \left ( b - z \right )
% \frac{w'}{w} - a = 0
% \label{eq:10}
% \end{equation}
% where $w(z)$ denotes the confluent hypergeometric function and $w'$
% denotes its derivative with respect to $z$, its final argument.  The
% quantities $a=1-if/2$ and $b=1$ are the first and second arguments of
% the function.  In particular this equation is satisfied where $w'/w=1$
% for any value of $z$, so ${\bar u}=v$ is one extremum.
% Furthermore, because the imaginary component of $w'/w$ is proportional
% to the derivative of the magnification, these are also extrema of
% the magnification.

% A second extremum comes from rewriting Eq.~(\ref{eq:10}) slightly and
% substituting in the values of $a$ and $b$ to yield
% \begin{equation}
% \frac{d}{dz} \left [ z \left ( \frac{w'}{w} -1 \right )\right ] + z
% \frac{w'}{w} \left (  \frac{w'}{w} - 1 \right )+ \frac{if}{2} = 0
% \end{equation}

Although several techniques exist to determine the range of the
confluent hypergeometric function \cite[e.g.][]{1632039}, it is
simpler in this case to resort to numerical experimentation,
it appears that the follow inequality obtains
\begin{equation}
0 \leq v \left ( {\bar u} - v \right ) \leq 2 f
\label{eq:11}
\end{equation}
with the value oscillating between the two extremes.  This yields useful
estimates for the magnitude of the astrometric shift from diffractive
lensing.  Furthermore, the value in the brackets of Eq.~\ref{eq:9} is
purely real at these extrema, so they are also extrema of the
magnification.

\subsection{Physical Optics} 

For values of $f, v \gg\ 1$ the value given by Eq.~(\ref{eq:8}) may be
estimated by using a physical optics approximation.  In particular the
square root of Eq.~(\ref{eq:1}) may be approximated up to a constant phase by
\citep{1992grle.book.....S}
\begin{equation}
I \approx e^{i\phi_+} \sqrt{\mu_+} + e^{i(\phi_--\pi/2)} \sqrt{\mu_-}
\label{eq:12}
\end{equation}
where
\begin{equation}
u_\pm = \frac{1}{2} \left ( v \pm \sqrt{v^2 + 4 f} \right ),
\end{equation}
\begin{equation}
\phi_\pm =  \frac{u_\pm^2}{2} - f\ln | u_\pm | - u_\pm v 
\end{equation}
and 
\begin{equation}
\mu_\pm = \frac{u_\pm}{v} \left |\frac{d u_\pm}{d v} \right |= \frac{1}{2} \left ( \frac{v^2+2 f}{v\sqrt{v^2+4 f}} \pm 1 \right ).
\end{equation}
The values $u_\pm$ are defined to be positive and negative while
$\mu_\pm$ is always positive.  The two images have opposite parity.
It is natural to interpret this as a negative value of the
magnification for one of the images; hence there is an additional
$\pi/2$ of geometric phase of the positive term relative to the
negative term.  This choice may seem rather arbitrary, but it results
from the stationary phase approximation of Eq.~(\ref{eq:1}) where one
extremum ($u_-$) is a saddle point.  This allows a simple expression like
Eq.~(\ref{eq:12}) to approximate the results of Eq.~(\ref{eq:1})
accurately for large values of $f$.  For more complicated lens
geometries, the phase lag is proportional to the Morse index of the
image \citep{1992grle.book.....S}.

Furthermore, essentially by design the following holds $d\phi_\pm/dv =
-u_\pm$.  Combining these results with Eq.~(\ref{eq:8}) yields an
estimate for
\begin{eqnarray}
{\bar  u} &\approx& \frac{1}{\mu} \Bigg \{ \mu_+ u_+ + \mu_- u_- +
\\
& & ~~~
\left (\mu_+ \mu_- \right )^{1/2} \bigg [ \left ( u_+ + u_- \right ) \cos \Delta \phi
\nonumber \\
& & ~~~  +  \frac{1}{2} \left ( \frac{d \ln \mu_+}{dv}
- \frac{d \ln \mu_+}{dv} \right ) \sin \Delta \phi \bigg ]
\Bigg \} \nonumber
\end{eqnarray}
where at this level of approximation the total magnification is
\begin{equation}
\mu \approx \frac{v^2 + 2 f \left ( 1 + \cos \Delta \phi \right
  )}{v\sqrt{v^2+4 f}}
\end{equation}
and
\begin{equation}
\Delta \phi = \phi_+ - \phi_- + \frac{\pi}{2} ~\mathrm{and}~ \mu_+ \mu_- =
\frac{f^2}{v^2 \left(v^2+4 f\right )}.
\end{equation}
The various definitions allow some further simplifications yielding
\begin{equation}
{\bar u} \approx \frac{1}{\mu} \left [
\frac{v^2 + f \left ( 3 + \cos\Delta \phi\right )}{\sqrt{v^2 + 4 f}}
+ \frac{2 f \sin \Delta \phi}{v \left (v^2 + 4 f\right)} \right ].
\label{eq:13}
\end{equation}
The result from geometric optics obtains by neglecting the terms with
$\Delta \phi$ yielding,
\begin{equation}
{\bar u} \approx \frac{v \left (v^2 + 3 f \right )}{v^2 + 2
  f}.
\label{eq:14}
\end{equation}
The maximum displacement due to lensing in the geometric limit is
$8^{-1/2} R_E$ at $v=2^{1/2} R_E$. 
\section{Results}
\label{sec:results}

\begin{figure}
\includegraphics[width=3.4in]{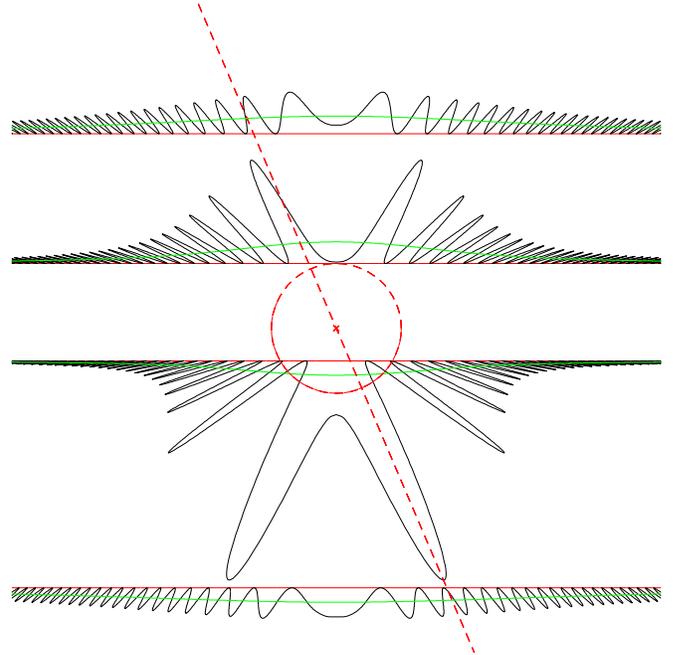}
\caption{The figure depicts the paths of images and sources for
  various impact parameters with the angular position of the lens
  fixed.  The dashed circle denotes the Einstein radius of the lens.
  The red line gives the path of the source (unlensed).  The green
  curve gives the path of the image without diffractive effects.  The
  black curve gives the path of the image with diffraction for
  ($f=10$).  The impact parameters from top to bottom are $3 R_E, R_E,
  0.5 R_E$ and $4 R_E$.  The dashed red line indicates how to
  translate a location along the path of the source to the location of
  the image centroid.}
 \label{fig:paths}
\end{figure}

Diffractive effects can have a dramatic effect on the trajectory of
images of gravitationally lensed sources.  In particular from
Fig~\ref{fig:paths} is it apparent that the maximal displacement is
much larger when diffractive effects are considered.  As in the
geometric limit, the centroid lies along the line connecting the
centre of the lens and the source.  Furthermore, the centroid lies
further from the centre of the lens than the source.  The observed
oscillations point back to the location of the lens, so the detection
of three oscillations combined with the presumably known proper motion
of the source determines the impact parameter between the source and
lens, the proper motion, mass and distance of the lens unequivocally.

\begin{figure}
\includegraphics[width=3.4in]{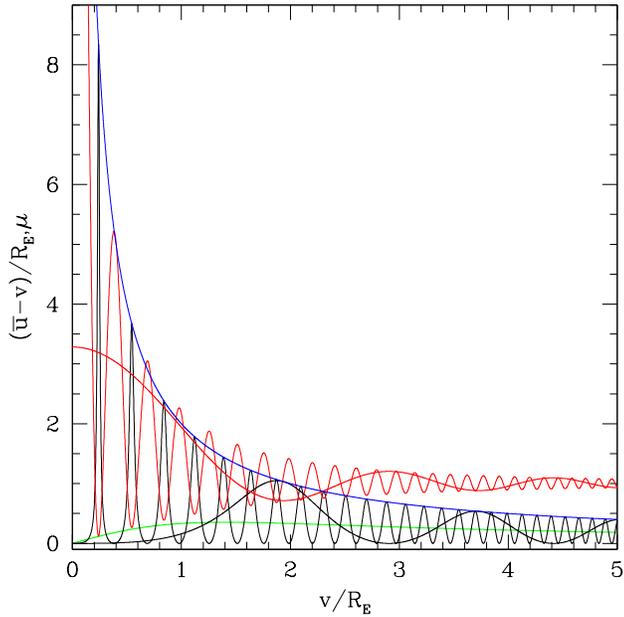}
\caption{The total magnification (red) and the difference (black)
  between the apparent radial position of the image centroid and the
  location of the source as a function of source location relative to
  the centre of the lens.  The green curve gives the displacement
  according to geometric optics.  The rapid oscillations are for $f=10$
as in Fig.~\ref{fig:paths}, and the slower oscillations are for
$f=1$.  The blue curve gives the envelope of $2 R_E^2/v.$}
\label{fig:posmag}
\end{figure}

Fig.~\ref{fig:posmag} shows that the displacement of the image
centroid from the source location oscillates between no displacement
and $2 R_E^2/v$ outward.  Furthermore, the minimal displacement occurs
at a maxima in the magnification.  The maxima of $v \left ({\bar u}
- v\right)$ (where the black curves touch the blue curve)
occur at a minima of the magnification.  In particular because for
small values of $v$ and large values of $f$ the magnification is well
approximated by a Bessel function \citep{1992grle.book.....S}, it is
straightforward to estimate the peak displacement that occurs near the
first zero of the Bessel function $J_0(x)$ at $x\approx 2.4$ to be
\begin{equation}
\left ({\bar u} - v\right)_\mathrm{max} \approx 0.83 f R_E ~\mathrm{at}~\frac{v}{R_E} \approx
2.4 f^{-1} ~\mathrm{for}~f\gg\ 1.
\label{eq:15}
\end{equation}
For smaller values of $f$, the peak displacement occurs for smaller
values of $v$ than given by this formula, and therefore the
displacement is larger than given here.  In particular, the
displacement is larger than the geometric limit (Eq.~\ref{eq:14}) for
$f>0.17$.  The maxima for $f=0.17$ occurs at $v/R_E \approx 5.5$ as
opposed to 14 as estimated from Eq.~(\ref{eq:15}).

\begin{figure}
\includegraphics[width=3.4in]{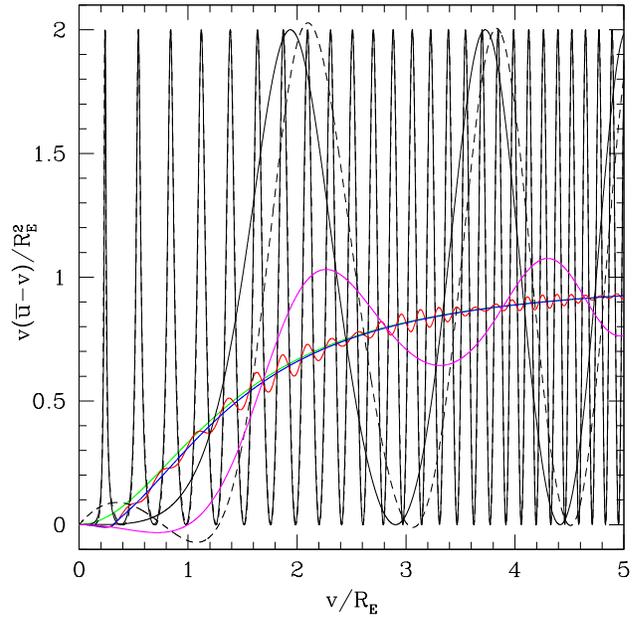}
\caption{The difference between the apparent radial position of the
  image centroid and the location of the source as a function of
  source location relative to the centre of the lens.  The result for
  $f=1$ is the slowly varying sinusoidal curve and $f=10$ is the more
  rapidly varying one.  The dashed curves give the results using the
  physical optics approximation from Eq.~(\ref{eq:13}).  For $f=10$ it
  is nearly indistinguishable from complete result.  The displacement
  from geometric optics is plotted in green.  Notice for $f=10$ there
  are about three peaks over a length of one Einstein radius for small
  values of $v$ and more for larger values.  The other colours assume
  that the angular radius of the source equals the Einstein radius.
  Blue is the geometric optics result, red is for $f=10$ and magenta
  is $f=1$.  }
 \label{fig:position}
\end{figure}

The envelope of the displacement, $2 R_E^2/v$, is robust regardless of
the value of $f$ or $v$, so it is natural to focus on the displacement
by dividing by the size of the envelope, yielding
Fig.~\ref{fig:position}.   The approximation from physical optics is
depicted by the dashed curve and follows the accurate calculation for $f=10$
very closely.   However, for $f=1$ the agreement is much poorer.
Furthermore,  the physical optics approximation does not precisely
follow the simple envelope as the deviations below zero and above two
manifest.  Even if one is more careful and approximates the
magnification as a Bessel function, the envelope only obtains
approximately.  The presence of the strict envelope results from an
apparently thus-far unknown property of the hypergeometric functions
and allows useful approximation of the possible signal.

Fig.~\ref{fig:position} also shows the effect of a finite source size
to wash out the observed oscillations toward the geometric optics
result.  It is not surprising that the rapid oscillations suffer a
greater decrement than for $f=1$.  However, observational realities
push the use of higher frequency observations to get finer angular
resolution.  For a fixed source size and impact parameter, the size of
the oscillations is proportional to $f^{-3/2}$ while the angular
resolution of a given telescope is proportional to frequency,
increasing with $f$.  Consequently if angular resolution is the only
factor in the accuracy of determining the centroid, the
signal-to-noise of the measurement of the centroid is proportional to
$f^{-1/2}$; it makes more sense to perform the measurement at lower
frequency.  On the other hand with increased flux, the centroid can be
determined more accurately, so this conclusion could change depending
on the spectrum of the object.

Determining the centroid of an object's emission is generally more
difficult than measuring the flux itself; therefore, searching for the
diffractive flux variation would generally be more fruitful than
looking for an astrometric signature, unless the flux from the source
is inherently noisy making the oscillation in the magnification
difficult to detect.  \citet{2010arXiv1002.3007H} outlines using
quasars as powerful tools to detect diffractive microlensing.  Flat
spectrum radio quasars generally have high brightness temperatures, so
large fluxes from small solid angles.  This can dramatically increase
the expected signal-to-noise ratio for a diffractive microlensing
event.  On the other hand the flux from the quasar may be inherently
noisy dominating the detector noise upon which
\citet{2010arXiv1002.3007H} focus.  For such objects the astrometric
signature of diffractive microlensing is a powerful tool.

Because of the envelope of the oscillation, it is straightforward to
estimate the magnitude of the displacement oscillation given the
impact parameter of the source relative to the lens ($b$) and the properties
of the lens itself.  In particular for a planetary mass lens the
following obtains
\begin{equation}
\left ( {\bar u} - v \right )_\mathrm{max} \approx
0.5 \frac{M_d}{\mathrm{M}_\oplus}  \left (
  \frac{D_d}{1~\mathrm{pc}} \right )^{-1} \!\!\!\!
\left ( \frac{b}{0.1~\mathrm{mas}}\right )^{-1} \!\!\!\! \mathrm{mas}
\label{eq:16}
\end{equation}
as along as the source may be considered compact compared to the
diffraction fringes.  In particular the SKA is expected to have an
angular resolution of about 5~mas at 20~GHz, or $f \approx 7$ for an
Earth-mass lens \citep{Schi07}.  Whether or not SKA measurements could
constrain the positions of quasars to less than a
milliarcsecond remains to be seen, but the VLBA typically measures the
positions of sources about 10~microarcseconds, so the SKA could in
principle achieve 60~microseconds with its larger minimum wavelength
and smaller size, detecting Earth-mass lenses out to about 10~pc with
a source impact parameter of 0.1~milliarcseconds.  If one were
especially lucky and found an especially close encounter between the
lens and source, the maximal displacement is
\begin{equation}
\left ( {\bar u} - v \right )_\mathrm{max} \approx 1
\left ( \frac{M_d}{\mathrm{M}_\oplus} \right)^{3/2} \!\!\! \left (
  \frac{D_d}{1~\mathrm{pc}} \right )^{-1/2} \!\!\!\!\!
  \frac{\nu}{20~\mathrm{GHz}} \mathrm{mas}
\label{eq:17}
\end{equation}
in principle detectable with the SKA out to 250~pc.  However, at such
a distance the lens subtends such a small angle that finite-source
effects are likely to be important.

\citet{2010arXiv1002.3007H} calculated the expected event rate of
substellar objects lensing bulge stars in the OGLE-II catalogue that
can also be detected with the Square Kilometre Array (about 80,000
stars) under the assumption that the density of substellar objects in
the disk of our Galaxy is about one-tenth of the total density.  The
total optical depth of such lenses is about $2\times 10^{-9}$. The
calculation neglected the possibility of substellar objects in the
Galactic halo and assumed that a lensing event lasts one day.  Under
these assumptions, the event rate where a source and lens align to
within one Einstein radius is about once per 14 years.  The lensing of
bulge stars could yield an astrometric signature of diffractive
microlensing in addition to the magnification signature discussed in
\citet{2010arXiv1002.3007H}.  

This event rate can be scaled to the microlensing rate for flat
spectrum radio quasars for which the astrometric signature may be
easier to measure.  To achieve a rate of once per decade about 100,000
radio sources would have to be monitored.  Because one expects the
lenses to be restricted to the plane of the Galaxy more or less, only
those quasars that lie within ten degrees of the Galactic equator
should be considered as sources (about two steradians).  The number
counts of flat-spectrum radio quasars give the expected event rate as
a function of the quasar radio flux.  \citet{2010A&ARv..18....1D}
compile the number of counts of radio-loud quasars at several
frequencies.  In particular above several GHz the sample will be
dominated by flat-spectrum sources; furthermore, this is where the SKA
is sensitive.  From the number counts at 8.4~GHz
\citep{1993ApJ...405..498W,2002AJ....123.2402F,2005ApJ...635..950H}
one can estimate that there are about $10^5$ flat-spectrum radio
sources within ten degrees of the Galactic plane with fluxes greater
than 4~mJy.  To get a sample of $10^6$ sources requires a flux limit
of 0.3~mJy.  What remains to be seen is how well future instruments
will be able to centroid such sources as a function of flux.  These
limits of course refer to the flux of the entire source not just the
high brightness temperature components that will show the most
dramatic astrometric signatures.

\section{Conclusions}
\label{sec:conclusions}

The continuous monitoring of compact, distant radio sources may
provide new way to probe the constituents of our solar neighbourhood, in
particular freely, floating sub-stellar objects.  The astrometric
signatures of diffractive microlensing can provide an estimate of the
mass, distance and proper motion of the lensing object, possibly
allowing follow-up observations of the lens itself.  Astrometric
lensing even without diffraction effects can provide this information
as well \citep{wambsganss06:_gravit_lensin}; however, diffraction
typically amplifies the astrometric signature and radio observations
often offer much higher angular resolution on the order of ten
milliarcseconds versus several hundred milliarcseconds in the optical.

This letter has used the specifications of the SKA as a benchmark.
Clearly the high angular resolution and high frequency offered by the
SKA are helpful for the detection of astrometric lensing in the radio;
however, the high sensitivity of the SKA may not strictly be necessary
if one focuses on bright radio sources.  Perhaps, a purpose-built
very-large baseline array of phased dipoles could achieve the needed
angular resolution (and possibly even a finer resolution than the SKA)
with a sufficient sensitivity to continuously determine the centroids
the brightest radio sources to the needed accuracy to detect low-mass
objects in the solar neighbourhood.  Furthermore, such a monitoring
campaign could yield new insights on quasar physics as well as other
ancillary results.  The low expected optical depth for these events of
about $2\times 10^{-9}$ would required the monitoring of 100,000 radio
sources to achieve even the modest event rate of once per decade.
These sources could be quasars or bulge giants, although the effect
should be more pronounced with the high brightness-temperture quasars.

\section*{Acknowledgments}

The Natural Sciences and Engineering Research Council of Canada,
Canadian Foundation for Innovation and the British Columbia Knowledge
Development Fund supported this work.  This research has made use of
NASA's Astrophysics Data System Bibliographic Services. 

\bibliographystyle{mn2e}
\bibliography{mine,physics,math,oort,paper2}
\label{lastpage}
\end{document}